# Creation, stabilization, and study at ambient pressure of pressure-induced superconductivity in $Bi_{0.5}Sb_{1.5}Te_3$

Liangzi Deng[1,8]*, Busheng Wang[2,8], Clayton Halbert[3], Daniel J. Schulze[1], Melissa Gooch[1], Trevor Bontke[1], Ting-Wei Kuo[1,4], Xin Shi[1], Shaowei Song[1], Nilesh Salke[5], Hung-Duen Yang[4], Zhifeng Ren[1], Russell J. Hemley[3,5,6], Eva Zurek[2], Rohit P. Prasankumar[7], Ching-Wu Chu[1]*

1. Department of Physics and Texas Center for Superconductivity at the University of Houston (TcSUH), Houston, Texas, 77204, USA

2. Department of Chemistry, University at Buffalo, Buffalo, New York 14260, USA.

3. Department of Chemistry, University of Illinois Chicago, Chicago, Illinois 60607, USA.

4. Department of Physics, National Sun Yet-Sen University, Kaohsiung 80424, Taiwan.

5. Department of Physics, University of Illinois Chicago, Chicago, Illinois 60607, USA.

6. Department of Earth and Environmental Sciences, University of Illinois Chicago, Chicago, Illinois 60607, USA.

7. Enterprise Science Fund, Intellectual Ventures, Bellevue, Washington 98005, USA

8. Co-first authors

* Liangzi Deng and Ching-Wu Chu

**Email:** ldeng2@central.uh.edu and cwchu@uh.edu

**Author Contributions:** L.Z.D., X.S., E.Z., and C.W.C. designed research; L.Z.D., B.S.W., C.H., D.J.S., M.G., T.B., T.W.K., X.S., S.W.S., N.S., R.J.H., and E.Z. performed research; H.D.Y. and Z.F.R. contributed new analytic tools; L.Z.D., R.P.P., and C.W.C. analyzed data; L.Z.D., C.W.C., B.S.W., and R.P.P. wrote the paper; L.Z.D. and C.W.C. directed the project; Z.F.R. supervised sample synthesis; E.Z. supervised theoretical research.

**Competing Interest Statement:** The authors declare no competing interests.

**Classification:** Physical Sciences: Physics

**Abstract**

In light of breakthroughs in superconductivity under high pressure, and considering that record critical temperatures ($T_c$s) across various systems have been achieved under high pressure, the primary challenge for higher $T_c$ should no longer solely be to increase $T_c$ under extreme conditions but also to reduce, or ideally eliminate, the need for applied pressure in retaining pressure-induced or -enhanced superconductivity. The topological semiconductor $Bi_{0.5}Sb_{1.5}Te_3$ (BST) was chosen to demonstrate our approach to addressing this challenge and exploring its intriguing physics. Under pressures up to ~ 50 GPa, three superconducting phases (BST-I, -II, and -III) were observed. A superconducting phase in BST-I appears at ~ 4 GPa, without a structural transition, suggesting the possible topological nature of this phase. Using the pressure-quench protocol (PQP) recently developed by us, we successfully retained this pressure-induced phase at ambient pressure and revealed the bulk nature of the state. Significantly, this demonstrates recovery of a pressure-quenched sample from a diamond anvil cell at room temperature with the pressure-induced phase retained at ambient pressure. Other superconducting phases were retained in BST-II and -III at ambient pressure and subjected to thermal and temporal stability testing. Superconductivity was also found in BST with $T_c$ up to 10.2 K, the record for this compound series. While PQP maintains superconducting phases in BST at ambient pressure, both depressurization and PQP enhance its $T_c$, possibly due to microstructures formed during these processes, offering an added avenue to raise $T_c$. These findings are supported by our density-functional theory calculations.

**Significance Statement**

As the renowned material scientist Pol Duwez once noted, almost all solids important to industry are in their metastable state. For example, all record high superconducting transition temperatures have been achieved only in materials under pressure, which are thus in a metastable state. Such extreme conditions present a significant obstacle to fully exploring these metastable phases and realizing the full promise of superconductivity. A solution is to stabilize them at ambient pressure. Here we used a pressure-quench protocol we recently developed to successfully stabilize the pressure-induced superconducting states in $Bi_{0.5}Sb_{1.5}Te_3$ at ambient pressure and up to room temperature. This achievement marks a significant step toward maintaining at ambient pressure novel metastable states generated under extreme conditions for science and technology.

**Main Text**

**Introduction**

All record-high superconducting critical temperatures ($T_c$s) reported to date occur in different materials under high pressures, *e.g.* 164 K in $HgBa_2Ca_2Cu_3O_{8+\delta}$ (Hg1223) at 31 GPa [1], 203 K in $H_3S$ at ~ 155 GPa [2], 260 K in $LaH_{10}$ at ~ 190 GPa [3], and most recently 80 K in $La_3Ni_2O_7$ above 14 GPa [4, 5]. This high-pressure condition poses serious challenges to the field of superconductivity, from science to applications. For example, many advanced characterization tools, such as scanning tunneling microscopy (STM), transmission electron microscopy (TEM), and angle-resolved photoemission spectroscopy (ARPES), are not accessible in high-pressure environments. Consequently, the physics of these novel states under high pressure remains largely unexplored. Furthermore, the intrinsic nature of certain critical states, such as superconductivity, still requires additional confirmation. One relief to this impasse is to find a way to retain the coherent quantum state under ambient conditions, *i.e.*, room temperature of ~ 300 K and ambient pressure. Our research has shown that the pressure-quench protocol (PQP) we developed [6, 7], *i.e.*, completely releasing the pressure ($P_Q$) at a specific temperature ($T_Q$) over a very short duration of time, enables the retention of high-pressure-induced/-enhanced superconducting phases in some compounds at ambient conditions.

For the present study, we chose the topological alloy compound $Bi_{0.5}Sb_{1.5}Te_3$ (BST) for a variety of reasons. In 2001, D. A. Polvani *et al.* attributed improved thermoelectric performance in BST to a possible Fermi surface topology change (FSTC) driven by high pressure, with a record room-temperature thermoelectric figure of merit (zT) of 2.2 at 1.7 GPa [8]. More than 20 years later in 2022, Chen's group and colleagues confirmed the pressure-enhanced zT in the same material [9]. FSTC has also been shown to play an important role in superconductivity [10-12], motivating us to use high-pressures to generate a superconducting state in the fully gapped bulk state of BST *via* a FSTC while keeping its chemistry and structure intact. For future rigorous investigation of high-pressure-induced metastable states in BST without interference from the high-pressure apparatus, *e.g.* the diamond anvil cell (DAC), we subsequently employed the PQP [6, 7] to retain such states at ambient pressure. Retaining possible topological superconducting states at ambient pressure could also provide an approach to tackle the longstanding open questions regarding chemically induced topological superconductivity, *e.g.* the debate regarding the topological nature of the superconducting compound $Cu_xBi_2Se_3$ [13-16].

Here we report our success in inducing superconductivity in three structural phases of BST (BST-I, -II, and -III) under different pressures and in retaining these phases at ambient pressure by PQP, in general agreement with our density-functional theory (DFT) calculations. Furthermore, we emphasize that successfully pressure-quenching a sample at low temperature and recovering it from the DAC at room temperature while retaining the pressure-induced phase at ambient pressure has profound and broad implications for exploring uncharted physics of metastable states and bridging the gap between research and applications. A possible topological superconducting state related to an electronic FSTC and without any structural change was detected in the BST-I phase under pressure. The pressure-quenched (PQed) sample not only maintained the high-pressure-induced superconducting state at ambient pressure, but it was also found to be a bulk superconductor. Employing the PQP above 30 GPa and at 77 K was also found to raise the $T_c$ to ~ 10.2 K, the highest to date for BST, at ambient pressure; this is presumably due to the microstructures generated during PQP. The pressure-induced superconductivity retained at ambient pressure and the reported pressure-enhanced thermoelectricity in this compound [8, 9] may also afford us the opportunity to search for the possible co-existence of superconductivity and thermoelectricity, an intriguing conjecture by Ginzburg that dates back to 1978 [17].

**Results and Discussion**

*Crystalline structures of BST under pressure*

The structures at room temperature of polycrystalline BST samples prepared by the standard solid state reaction method (details are provided in the Materials and Methods section) were examined under pressure up to 54 GPa using the Advanced Power Source at Argonne National Laboratory. Selected X-ray diffraction (XRD) patterns are presented in Fig. 1. Three phases, BST-I, -II, and -III, were observed in different pressure regions. Mixed phases also exist near the phase boundaries, as shown in Fig. S1, indicating a high reorganization energy or kinetic barrier between each set of bordering phases. This may be at least partially responsible for our success in retaining these pressure-induced phases at ambient pressure and is consistent with our calculations. Near ambient conditions, BST crystallizes in an $R\bar{3}m$ rhombohedral structure (BST-I) with lattice parameters, *e.g.* at 1.1 GPa, of *a* = 4.2640(4) Å, *c* = 30.084(4) Å (Z = 3). As the pressure increases to above 10 GPa, a structural phase transition occurs, in which the compound enters a monoclinic *C*2/*m* structure (BST-II) with lattice parameters, *e.g.* at 11.0 GPa, of *a* = 14.327(7) Å, *b* = 4.1544(3) Å, *c* = 17.849(8) Å, β = 149.64(1)° (Z = 4). Above 23 GPa, an $Im\bar{3}m$ body-centered cubic phase (BST-III) appears with lattice parameter *a*, *e.g.* at 23.4 GPa, of 3.5252(5) Å (Z = 2). These high-pressure results compare well with the calculated XRD data, as shown in Fig. 1c. The bulk modulus $K_0$ obtained from the X-ray diffraction measurements is 35 GPa, which corresponds to a compressibility $\beta_0$ of 0.029 GPa$^{-1}$.

*Pressure-dependent room-temperature resistance of BST*

At ambient pressure, the resistivity (ρ) of BST decreases smoothly with decreasing temperature (T) down to 2 K, as shown in the inset to Fig. 2. This metallic behavior, in contrast to the predicted semiconducting bulk, may be associated with the metallic surface state of the compound. With increasing pressure, the pressure-dependent resistance R(P) at room temperature decreases rapidly up to ~ 10 GPa (gray area in Fig. 2), and slowly above, but levels off above ~ 24 GPa (red arrow in Fig. 2). These room-temperature R(P) anomalies take place at pressures where structure transformations occur based on the XRD results shown in Fig. 1, *i.e.*, from BST-I to BST-II and from BST-II to BST-III at ~ 10 and ~ 23 GPa, respectively. A weak anomaly near ~ 39 GPa, where no structural change appears in the XRD results, was also detected, as marked by a red arrow in Fig. 2, and is possible evidence for an additional FSTC.

*Superconductivity in BST under pressure*

BST becomes superconducting under pressure above 3.3 GPa with an onset $T_c$ (defined in Fig. 3a and referred to as $T_c$ henceforth) up to ~ 9 K. R(T) curves for BST at temperatures up to 12 K under pressures up to ~ 45.5 GPa during pressurization and during depressurization are shown in Figs. 3a and 3b, respectively. We would like to note that, as shown in Fig. 3a, the superconducting transition with a $T_c$ ~ 4.8 K appears suddenly at 4.8 GPa in BST-I, in which no sign of crystal symmetry changes or structural transition is evident below 10 GPa (Fig. 1). A detailed lattice *c/a* ratio analysis does show an anomaly at 4-5 GPa (inset, Fig. 4), similar to what has been previously reported for the parent compound $Bi_2Te_3$ [18]. Distinct changes in Fermi surface (FS) topology above 4 GPa are also revealed by our DFT calculations (Fig. S2). All these observations suggest that a FSTC has taken place under pressure. Since BST-I is topologically nontrivial, one may reasonably assume that the pressure-induced superconducting BST-I is a topological superconductor under ~ 4.8 – 10 GPa, similar to what was suggested by Jin's group [19] for the parent compound $Bi_2Te_3$ with a $T_c$ at 3 K. Further increasing pressure leads to a continuous increase of $T_c$ up to 5.1 K at 8.1 GPa. An abrupt $T_c$ jump is evident near 10 GPa, coinciding with the occurrence of the BST-I → BST-II crystal structural transition. $T_c$ continues to increase with increasing pressure up to ~ 16 GPa, peaking at ~ 9 K before beginning to decrease. A weak $T_c$ anomaly appears at ~ 24 GPa, coinciding with the BST-II–BST-III boundary. $T_c$ continues to decrease with a further increase in pressure but starts to increase at ~ 35 GPa and exhibits a small peak at ~ 39 GPa, where no structural transition is evident from the XRD data (Fig. 1). The unusual hysteretic variations of $T_c$ with pressure between pressurization and depressurization are summarized in Fig. 4.

*Retention of the high-pressure superconducting BST phases at ambient pressure by PQP*

BST clearly displays different unusual quantum states under pressure. Unraveling the detailed physics of these states will be most rewarding. This will require carrying out rigorous characterization of these states without the influence of the DAC. We have therefore examined their retainability at ambient pressure by employing the PQP recently developed by us [6, 7]. We first examined the hysteretic effect between pressurization and depressurization (Fig. 4). As mentioned earlier, the XRD results in Fig. 1 indicate a high reorganization energy or kinetic barrier between each set of bordering phases. This is evident from the R(P) results obtained during slow depressurization at room temperature, which normally takes a few minutes or longer (Fig. 3b). In other words, pressure hysteresis exists for the phase transitions upon the slow withdrawal of pressure. In addition, we observed that the $T_c$ of the depressurized sample becomes up to 23% higher than that of the pressurized one at the same pressure for BST-II and -III and even higher for BST-I. Although slow depressurization to 1.2 GPa can retain superconductivity with a $T_c$ ~ 7 K, this method fails to retain the pressure-induced superconductivity, or even the low-resistivity state, when pressure is completely removed, as shown in Figs. 3b and 4.

On the other hand, by employing the PQP, we have successfully retained at ambient pressure the pressure-induced superconducting phases, as shown in Figs. 5a-c for samples PQed at different quenching pressures ($P_Q$) at 77 K and in Fig. 5d for a sample PQed at 32.3 GPa at 4.2 K. In Figs. 5a-d, we have also included the R(T) curves at chosen quenching pressures $P_Q$s for specific phases realized before PQ (*i.e.*, rapid removal of the pressure to ambient pressure) at a specific quenching temperature $T_Q$. The R(T) results under 8.0 GPa in Fig. 5a (■) show the possible topological superconducting transition in BST-I with a $T_c$ ~ 4.8 K. The small resistance peak right before the superconducting transition may result from some yet-to-be-determined pressure-induced microstructures. This potential pressure-induced topological superconducting phase is clearly retained at ambient pressure after PQ at 8.0 GPa (and other $P_Q$s) and 77 K (and other $T_Q$s) *via* PQP, as shown, *e.g.*, by the ambient-pressure R(T) results in Fig. 5a (●). The high-pressure R(T) results in Fig. 5b (■) show that, under 20.3 GPa, the sample is in its BST-II phase with a $T_c$ of ~ 8.3 K. However, after PQ at 20.3 GPa and 77 K, only the ~ 4.9 K superconducting phase can be retained at ambient pressure (Fig. 5b, ●).

The R(T) results under 33.3 GPa in Fig. 5c (■) show that the sample is in the BST-III phase with a $T_c$ of 6.2 K. After being PQed at 33.3 GPa and 77 K, a mixed phase was retained at ambient pressure with two different $T_c$s at ~ 6 K and ~ 10.2 K (Fig. 5c, ●). The superconducting nature of the 10.2 K transition was further demonstrated by current effect testing (Fig. S3). It should be noted that $T_c$ of 10.2 K is the highest reported for this compound series under pressure; $T_c$s of up to 8 K in Bi at 8.1 GPa [20], 4.8 K in Sb at 24 GPa [7], 8.1 K in $Bi_2Te_3$ at 8.9 GPa [21-23], 7.3 K in $Sb_2Te_3$ at 30.5 GPa [24], and 8.9 K in $Bi_2Te_{2.1}Se_{0.9}$ at 18.3 GPa [25] have been previously reported. Based on the $T_c$s of the phases retained at ambient pressure by PQP, it is clear that these phases are the possible topological superconducting phase with a $T_c$ ~ 5 K in BST-I and a different superconducting state with a record high $T_c$ above 10 K in BST-II or -III or even a phase different from BST-I, -II, or -III. It has been shown that lower $T_Q$ can help retain the metastable phases with higher $T_c$ for FeSe [13]. We therefore PQed BST at $T_Q$ = 4.2 K. Unfortunately, only the superconducting phase with a $T_c$ ~ 5.9 K could be stabilized at ambient pressure (Fig. 5d, ●).

The above observations show that the possible topological superconducting state with a $T_c$ ~ 5 K in BST-I induced by ~ 4.8 GPa < P < 10 GPa and the superconducting state with a $T_c$ up to ~ 10.2 K in BST-II or -III or a new phase of BST induced by P > 10 GPa are retained at ambient pressure by PQP for the future unraveling of the intriguing properties predicted for these states.

We would like to point out that the PQP is only in its initial stages of refinement, and additional systematic studies are being developed to allow the PQP to live up to its full potential, *e.g.,* controlling the dP/dt; determining the structure evolution details during PQP and the stability of the PQed phases; and establishing the protocol for characterizing the PQed phases, especially their microstructure *via* spectroscopy measurements after PQed sample recovery from the DAC.

### *Thermal and temporal stability of the PQed BST phases*

The practicality of these PQed phases for science and application studies will be determined by their thermal and temporal stability. We therefore examined the stability of the PQed phases with variations in temperature and time. The effects on the superconducting transitions of PQed phases by thermal excursions from 20 K sequentially up to 300 K are shown in Fig. S4 by the ambient-pressure ρ(T) results of the possible topological superconducting phase retained in BST after being PQed at $P_Q$ = 31.6 GPa and $T_Q$ = 77 K. As shown in Fig. S4, the higher superconducting transition with $T_c$ ~ 10 K starts to degrade when the excursion temperature rises to $T_Q$ ~ 77 K (curves 4-8). However, even when the excursion temperature is raised to 300 K, a substantial part of the superconducting transition survives (Fig. S4a). The lower onset of the transition with $T_c$ ~ 6.0 K remains unaffected when the excursion temperature is kept below 150 K. The ρ(T) of a sample PQed at 33.3 GPa and 77 K was additionally measured repeatedly between 4.2 and 77 K for a period of 5 days (Fig. S5). The higher superconducting phase with $T_c$ ~ 10.2 K appeared to have slightly degraded over the long-term testing while the lower $T_c$ ~ 6.0 K showed negligible change. However, at the same time, small temperature increases in both superconducting transitions were detected, as indicated by the dashed arrows in Fig. S5.

*The bulk nature of the PQed superconducting phase in BST*

Since the PQed superconducting phase in BST is at least partially stable up to 300 K as shown in Fig. S4a, we decided to determine if this phase is bulk or filamentary. The R(T) and DC magnetic susceptibility χ(T) = M(T)/H of a sample ~ 10 μm thick and ~ 65 μm across PQed at 33.3 GPa and 77 K and recovered from the DAC (inset, Fig. 5f) were measured, and the results are shown in Figs. 5e-f, respectively. As shown in Fig. 5f, a diamagnetic shift corresponding to ~ 46% of the superconducting transition below ~ 4 K was detected, confirming the bulk nature of the PQed superconducting state. The superconducting transition at ~ 10 K shown by the R(T) results in Fig. 5c and by the ρ(T) results in Figs. S4 and S5 did not survive the large thermal and temporal variations experienced by the sample during its handling, while the superconducting transition at 4-6 K does fortunately survive, as also evidenced from Fig. 5e-f. Significantly, this shows that a pressure-induced superconducting phase has not only been successfully retained at ambient pressure but could also be recovered from the DAC at room temperature while still maintaining its superconducting properties at lower temperatures. These observations demonstrated the prowess of PQP for the rigorous investigation of metastable phases induced under high pressure without the interference of high-pressure equipment like DACs, provided special care is taken concerning the thermal and temporal stability of the PQed phase. Future studies will be performed on the PQed BST samples, such as X-ray diffraction, transmission electron microscopy (TEM), and angle-resolved photoemission spectroscopy (ARPES), to determine the retained crystal phase and microstructure, as well as its topological properties.

*DFT analyses of metastability and retainability of high-pressure BST phases*

To obtain a better understanding of the crystallinity of BST under various pressures and the lattice dynamics of these phases after depressurization, we performed DFT calculations. The calculated phonon dispersions shown in Figs. 6a-c and 6f respectively confirmed the dynamic stability of the following BST phases at the given pressures, as evidenced by the absence of imaginary frequencies: $R\bar{3}m$ (BST-I) at 0 GPa, $C2/m$ (BST-II) at 10 GPa and 20 GPa, and $Im\bar{3}m$ (BST-III) at 40 GPa. Phonon dispersions were also calculated to explore the stability of these phases upon depressurization, and it was found that BST-II survives down to 1 GPa (Fig. 6d) but not to 0 GPa (Fig.6e, which shows the occurrence of negative dispersion). BST-III can be retained down to 15 GPa (Fig. 6g) but not to 0 GPa (Fig. 6h). The downward shift in the stability range of BST-II to 1 GPa suggests the increased possibility for success in employing PQP to retain this phase at ambient pressure, in agreement with the experimental results (Figs. 4-5). Notably, our PQP results showed that superconductivity can be retained at ambient pressure (Fig. 5), whereas none of the high-pressure phases were predicted to be local minima within the harmonic approximation. However, anharmonic or finite-temperature effects on the lattice dynamics are known to renormalize phonon frequencies, thereby altering the range of dynamic stability [25-27]. The difference observed between regular depressurization and PQP is worth further theoretical analysis.

To explore qualitatively the electronic transport properties of BST during pressurization and depressurization processes, the electronic energy spectra were calculated for BST-I, -II, and -III, as shown in Fig. S6. BST-I is clearly a bulk insulator with a narrow indirect bandgap of 63 meV at 0 GPa that transforms into a narrow direct bandgap of 15 meV at 10 GPa. The metallic behavior observed experimentally (inset, Fig. 2) is likely related to the presence of the small energy gap near the Γ-point due to band inversion (Fig. S7). To account for the doping system in BST, we utilized the virtual crystal approximation in Quantum Espresso, which posed a technical challenge for calculating the properties of the surface states. However, considering that $Bi_2Te_3$, which has been identified as a topological material [3,4,19], and BST-I share the same space group ($R\bar{3}m$), along with the presence of a band gap and band inversion in $R\bar{3}m$, as shown in Fig. S5, it is natural to deduce that BST-I holds promise as a topological material. Above 10 GPa, BST becomes metallic and also superconducting, as observed experimentally. Both of the high-pressure phases of BST, $C2/m$ and $Im\bar{3}m$, exhibit metallic characteristics at high pressure and after depressurization (Figs. S6c-6f), suggesting that the persistence of superconductivity at lower pressures following regular depressurization or at ambient pressure following PQP could potentially arise from these phases. To establish a potential relationship between superconductivity and electronic properties across diverse pressure levels, we have plotted the density of states (DOS) values at the Fermi level ($E_F$) for different phases of BST as a function of pressure (Fig. S8). The DOS values at $E_F$ for BST-II and -III show general trends with pressure (Fig. S8), *i.e.*, the DOS for BST-II increases rapidly with increasing pressure and that for BST-III decreases slowly with increasing pressure, which is in general agreement with the experimental $T_c$ *vs.* P results (Fig. 4). Interestingly, the DOS value at ~ 20 GPa during depressurization is higher than that during pressurization, which is consistent with the discovery of depressurization-enhanced $T_c$ (Fig. 4). The calculated $T_c$s under different pressures during pressurization and depressurization (Table S1) are also qualitatively consistent with our experimental results.

**Conclusions**

In summary, we have conducted a comprehensive study of the structure, transport, and magnetic properties of the topological solid $Bi_{0.5}Sb_{1.5}Te_3$ at ambient conditions and under high pressures. Three distinct crystal structures were identified up to 53 GPa: BST-I (rhombohedral, $R\bar{3}m$) up to 10 GPa, BST-II (monoclinic, $C2/m$) between 10 and 23 GPa, and BST-III (body-centered cubic, $Im\bar{3}m$) between 23 and 53 GPa, in general agreement with our DFT calculations of the phonon and electronic energy spectra. An unusual superconducting transition was detected in the BST-I phase in the absence of any crystal symmetry change but coinciding with the appearance of a lattice parameter ($c/a$) anomaly characteristic of a FSTC. Our DFT analysis provides further evidence for FSTC above 4 GPa. We therefore propose that this superconducting state with a $T_c$ ~ 4.8 K arising from an electronic transition or FSTC in a topological insulator under 4-5 GPa may be the long-sought-after topological superconducting state, with significant implications for the development of the emerging quantum technology. We have successfully used our PQP to retain this unusual, possibly topological, superconducting state at ambient pressure outside a DAC and demonstrated its bulk nature, making feasible further detailed characterization on metastable states previously only maintained under pressure *via* heretofore inaccessible techniques such as scanning tunneling microscopy (STM), angle-resolved photoemission spectroscopy (ARPES), *etc.* Both thermal and temporal tests demonstrated that the superconducting state in BST-I exhibits rather high stability. We have also found $T_c$ enhancement through the processes of depressurization and PQP, offering an added novel avenue for raising $T_c$. Our results suggest that BST, with its superior superconducting and thermoelectric properties, may serve as an excellent platform to search for the possible coexistence of superconductivity and thermoelectricity as first suggested by Ginzburg. The potential impacts of the PQP and of the PQed phases retained at ambient pressure for unraveling the physics of pressure-induced metastable phases—not limited to superconductivity nor restricted to states generated by structural phase transitions—are therefore evident and could spur a host of new scientific directions across physics, chemistry, and materials science.

**Materials and Methods**

*Sample preparation*
The polycrystalline samples of $Bi_{0.5}Sb_{1.5}Te_3$ investigated were prepared by the standard solid-state reaction method. Stoichiometric amounts of bismuth (Bi, Alfa Aesar, 99.99%, pieces), antimony (Sb, Alfa Aesar, 99.8%, shots), and tellurium (Te, Alfa Aesar, 99.999%, pieces) were weighed according to the stoichiometry of $Bi_{0.5}Sb_{1.5}Te_3$ and loaded into an evacuated quartz tube that was subsequently heated at ~ 1023 K for 5 hours. The ingot obtained was pulverized in a high-energy ball-mill (SPEX 8000X). Finally, the powder was loaded into a graphite die and hot-pressed to ~ 60 MPa at ~ 773 K for 5 minutes.

*X-ray measurements under pressure*
Diamond anvil cells (DACs), each with 300-micron culets, were used to prepare samples for high-pressure measurement. Each rhenium gasket was preindented to 35-micron thickness and a 180-μm hole was drilled. Neon was used as the pressure-transmitting medium and ruby and gold were used as pressure markers. The DACs were prepared in an inert atmosphere. High-pressure X-ray diffraction measurements were carried out at beamline 16IDB sector 16, HPCAT, Advanced Photon Source (APS), Argonne National Laboratory (ANL), using a wavelength of 0.4066 Å. Diffraction images were integrated using Dioptas [28]. GSAS-II [29] and Jana [30] software packages were used to refine the lattice parameters by Le Bail refinement. Pressure-volume data were fitted using Vinet equations of state in the EosFit-7c [31] and EosFit GUI [32] software packages. Crystal structures were created using VESTA (Visualization for Electronic and Structural Analysis) [33].

*Electrical transport measurements under pressure*
DACs with various culets (300 and 400 μm) were used for high-pressure transport measurements. Electrodes were prepared using 5-μm-thick platinum foil and were insulated from the rhenium gasket in each DAC with a c-BN/epoxy mixture. A 100 – 130 μm diameter hole was drilled into the center of each gasket to serve as the sample chamber. A piece of $Bi_{0.5}Sb_{1.5}Te_3$ with a diameter of 80 – 100 μm and thickness of 10 – 20 um was loaded into the sample chamber with NaCl serving as the pressure-transmitting medium (PTM). The pressure was determined using the ruby fluorescence scale [34] or the diamond Raman scale at room temperature [35]. The contacts for each sample were arranged in a Van der Pauw configuration and data were collected using a Quantum Design Physical Property Measurement System (PPMS) or a Keithley 6221/2182A Delta Mode System.

*Magnetization measurements*
Magnetization measurements from 10 K to 2 K under 50 Oe were carried out using a Quantum Design Magnetic Property Measurement System (MPMS).

*Density-functional theory calculation*
All density-functional theory (DFT) and density-functional perturbation theory (DFPT) calculations of electronic and vibrational properties were carried out using the plane-wave pseudopotential code QUANTUM ESPRESSO [36], scalar-relativistic optimized norm-conserving Vanderbilt pseudopotentials (ONCV) [37], and the PBE-GGA exchange and correlation functional [38]. We used an 8×8×8 Monkhorst-Pack k-grid, a 4×4×4 Monkhorst-Pack q-grid, a kinetic energy cutoff of 100 Ry, a smearing of 0.06 Ry, an electronic convergence threshold of $10^{-10}$ Ry, and a phonon self-convergency threshold of $10^{-16}$. The relaxation thresholds for the BFGS steps were $10^{-7}$ Ry in total energy and $10^{-6}$ Ry/bohr for all force components. For all structures, we utilized the primitive unit cell consisting of 5 atoms for $R\bar{3}m$, and 10 atoms for each of the $C2/m$ and $Im\bar{3}m$ phases. Doping was simulated using the virtual crystal approximation (VCA) function within QUANTUM ESPRESSO, which amounts to replacing each Sb (pseudo)atom in the $Bi_2Te_3$ structure with an average virtual (pseudo)atom, obtained by mixing Sb and Bi in the appropriate proportions. It should be noted that the experimentally observed $Im\bar{3}m$ symmetry can only be assigned to this structure if the Sb-Bi and Te atoms are not distinguished. Treating the Sb/Bi atoms *via* the VCA, but differentiating the Te atom, reduces the symmetry of BST-II to $C2/m$ in our calculations. The critical superconducting temperature, $T_c$, has been estimated using the Allen-Dynes modified McMillan equation [39] as,

$$T_c = \frac{\omega_{log}}{1.2} exp\left[-\frac{1.04(1+\lambda)}{\lambda - \mu^*(1+0.62\lambda)}\right]$$

where $\omega_{log}$ is the logarithmic average frequency, $\lambda$ is the electron phonon coupling parameter, and $\mu*$ is the Coulomb pseudopotential, often assumed to be between ~0.1 - 0.16; a value of 0.1 was employed for the $R\bar{3}m$ and $C2/m$ phases and a value of 0.16 was employed for the $Im\bar{3}m$ phase.

**Acknowledgments**

We would like to thank P. C. Dai at Rice University for insightful discussions and suggestions. L. Z. D., D. J. S., M. G., T. B., T. W. K., and C. W. C. are supported by the Enterprise Science Fund of Intellectual Ventures Management, LLC; U.S. Air Force Office of Scientific Research Grants FA9550-15-1-0236 and FA9550-20-1-0068; the T. L. L. Temple Foundation; the John J and Rebecca Moores Endowment; and the State of Texas through the Texas Center for Superconductivity at the University of Houston. B. W. and R. J. H. acknowledge funding from the U.S. Department of Energy, Office of Science, Fusion Energy Sciences funding the award entitled High Energy Density Quantum Matter under Award No. DE-SC0020340. Calculations were performed at the Center for Computational Research at SUNY Buffalo (https://hdl.handle.net/10477/79221). C. H., N. S., R. J. H., and E. Z. are supported by the U.S. National Science Foundation (NSF) grants DMR-2119308 and DMR-2118020, and by the U.S. Department of Energy-National Nuclear Security Administration (DOE-NNSA) cooperative agreement DE-NA-0003975 (Chicago/DOE Alliance Center, CDAC). E. Z. is also supported by the U.S. National Science Foundation (NSF) grant DMR-2119308. Synchrotron X-ray diffraction experiments were performed at HPCAT (Sector 16), Advanced Photon Source (APS), Argonne National Laboratory (ANL). HPCAT operations are supported by DOE-NNSA's Office of Experimental Sciences. The APS is a DOE Office of Science User Facility operated for the DOE Office of Science by ANL under contract DE-AC02-06CH11357. R. P. P. is supported by the Enterprise Science Fund of Intellectual Ventures Management, LLC. We are grateful to Yue Meng for help with the X-ray measurements.

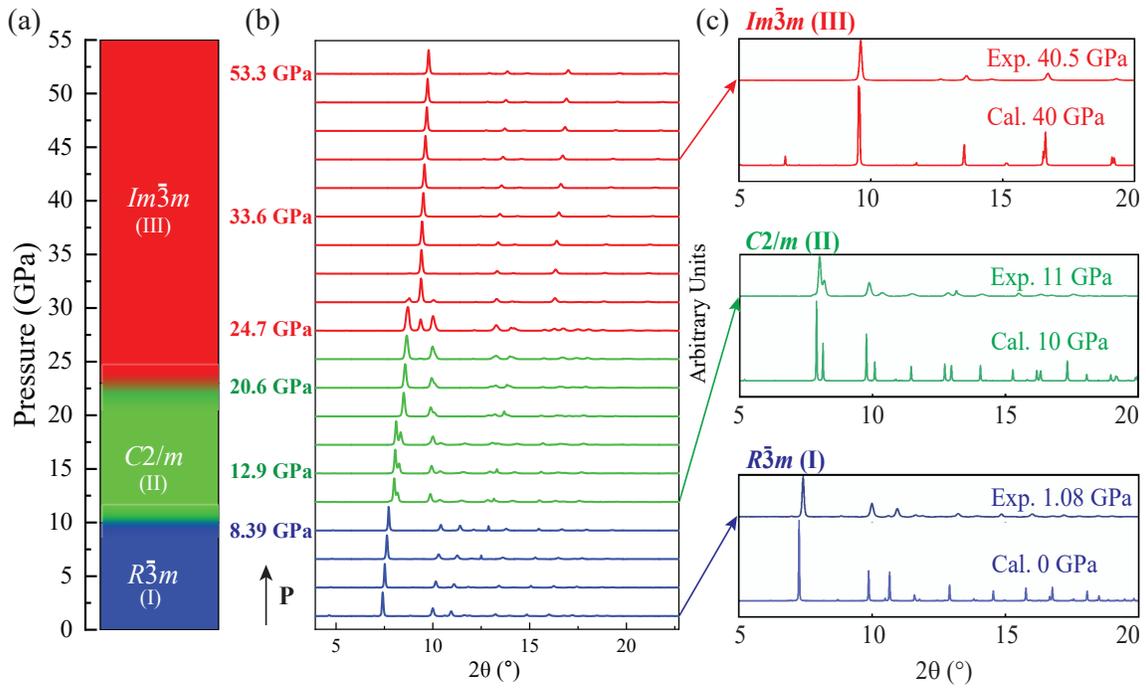

**Figure 1.** (a) Pressure-dependent crystal structure evolution of $Bi_{0.5}Sb_{1.5}Te_3$. With increasing pressure, its structure transitions from $R\bar{3}m$ (BST-I, blue) to $C2/m$ (BST-2, green) to $Im\bar{3}m$ (BST-3, red). (b) High-pressure X-ray diffraction (XRD) measurements at Argonne National Laboratory (ANL), using a wavelength of 0.4066 Å. The patterns around 24 GPa exhibit peaks from both the $C2/m$ and $Im\bar{3}m$ phases, indicating phase coexistence. (c) Comparison of selected experimental (Exp.) and calculated (Cal.) XRD results for $R\bar{3}m$, $C2/m$, and $Im\bar{3}m$ at 0, 10, and 40 GPa, respectively. The calculated XRD patterns were obtained for DFT-optimized structures.

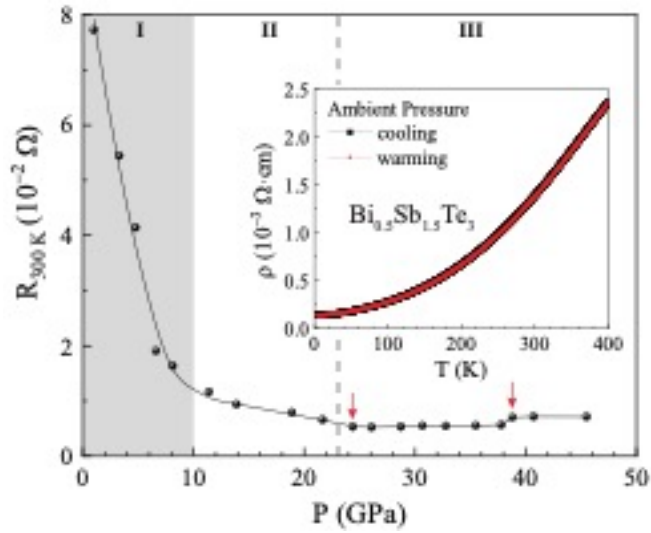

**Figure 2.** Room-temperature resistance of $Bi_{0.5}Sb_{1.5}Te_3$ as a function of pressure. The gray area represents the pressure range (P < 10 GPa) in which the resistance drops rapidly under pressure at a relatively high dR/dP rate. With increasing pressure, the resistance levels off above ~ 24 GPa and a weak anomaly appears near 39 GPa (red arrows). Inset: resistivity of $Bi_{0.5}Sb_{1.5}Te_3$ as a function of temperature from 2 K to 400 K measured under both cooling (■) and warming (●) at ambient pressure.

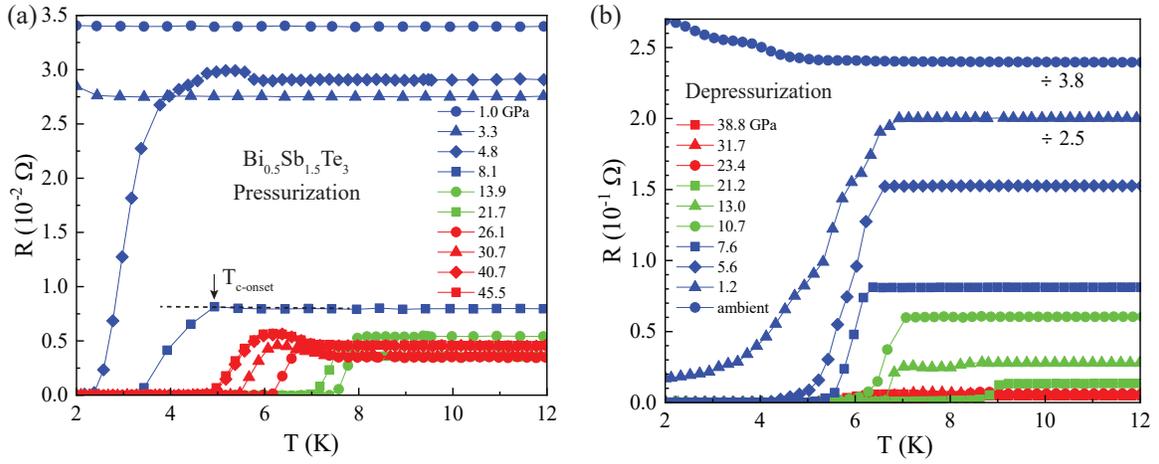

**Figure 3.** Resistance as a function of temperature for $Bi_{0.5}Sb_{1.5}Te_3$ under different pressures (a) from 1.0 to 45.5 GPa during pressurization and (b) from 38.8 GPa to ambient pressure during depressurization. The resistance values under 1.2 GPa and at ambient pressure were reduced by factors of 2.5 and 3.8, respectively, to fit the scale. Blue, green, and red symbols represent BST-I, -II, and -III, respectively.

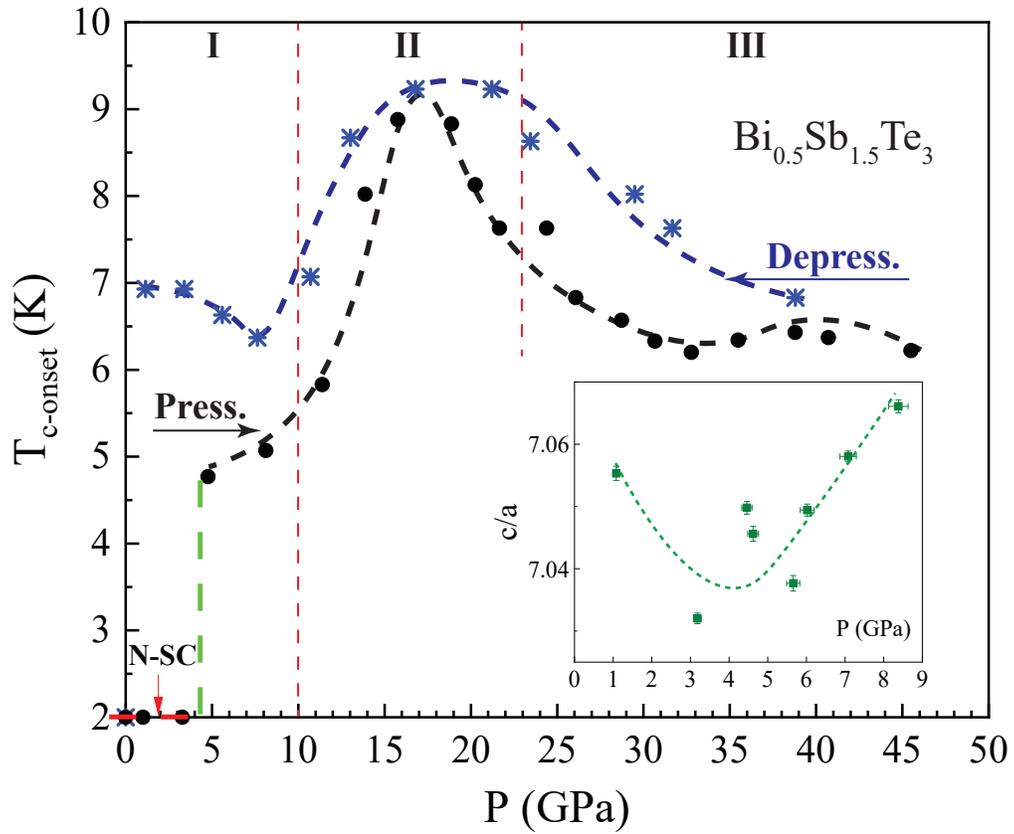

**Figure 4.** $T_{c\text{-onset}}$ as a function of pressure up to 45.5 GPa for $Bi_{0.5}Sb_{1.5}Te_3$ during pressurization ("Press.", ●) and depressurization ("Depress.", ✱). "N-SC" stands for non-superconducting. The green dashed line marks the critical pressure for the electronic topological transition. Red dashed lines indicate the critical pressures for structural phase transitions determined by synchrotron X-ray measurements. Inset: lattice *c/a* ratio as a function of pressure for $Bi_{0.5}Sb_{1.5}Te_3$ based on X-ray data obtained at Advanced Photon Source (APS, ■). Dashed curves in the figure and its inset are guides for the eye.

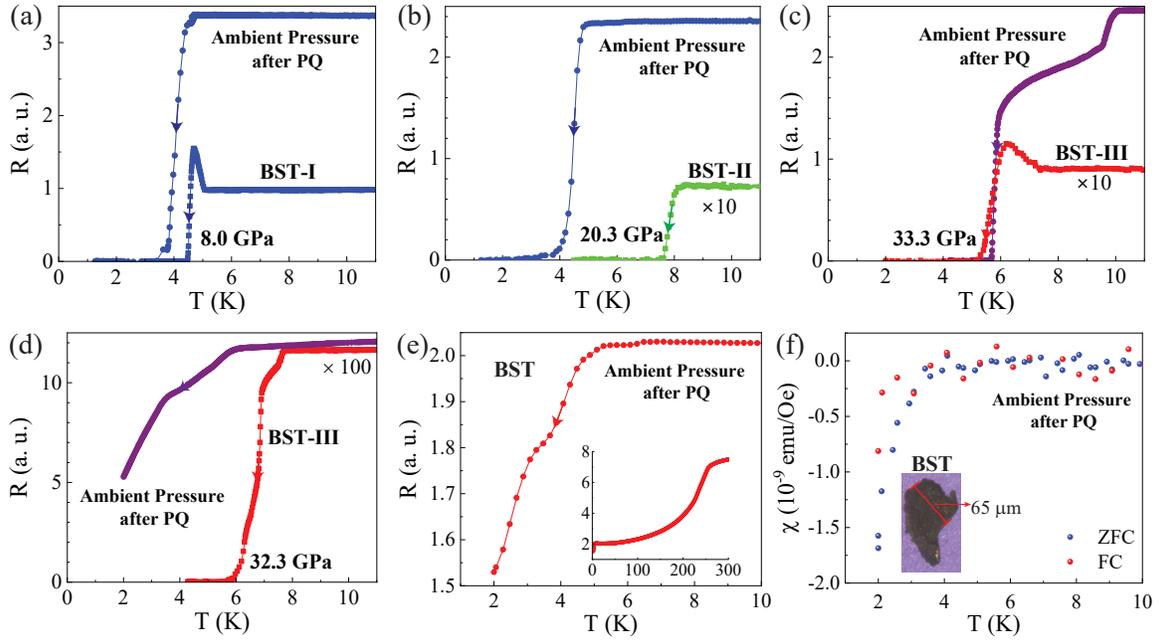

**Figure 5.** Characterization of Bi$_{0.5}$Sb$_{1.5}$Te$_3$ under pressure P and at ambient pressure after pressure quench (PQ) at quenching temperature T$_Q$ from quenching pressure P$_Q$. Temperature-dependent resistance [R(T)] under P (squares) and at ambient pressure after PQ (circles) for (a) P = 8.0 GPa, T$_Q$ = 77 K, P$_Q$ = 8.0 GPa; (b) P = 20.3 GPa, T$_Q$ = 77 K, P$_Q$ = 20.3 GPa [R(T) results under 20.3 GPa multiplied by 10]; (c) P = 33.3 GPa, T$_Q$ = 77 K, P$_Q$ = 33.3 GPa [R(T) results under 33.3 GPa multiplied by 10]; and (d) P = 32.3 GPa, T$_Q$ = 4.2 K, P$_Q$ = 32.3 GPa [R(T) results under 32.3 GPa multiplied by 100]. (e) R(T) at ambient pressure for a sample PQed at T$_Q$ = 77 K and P$_Q$ = 33.3 GPa and recovered from the diamond anvil cell. Inset: R(T) from 300 K to 2 K. Arrows in (a-e) indicate that all data were taken during cooling. (f) Susceptibility of the sample recovered from the diamond anvil cell as a function of temperature up to 10 K measured under both zero-field-cooled (ZFC) and field-cooled (FC) conditions. Inset: photograph of a recovered PQed sample.

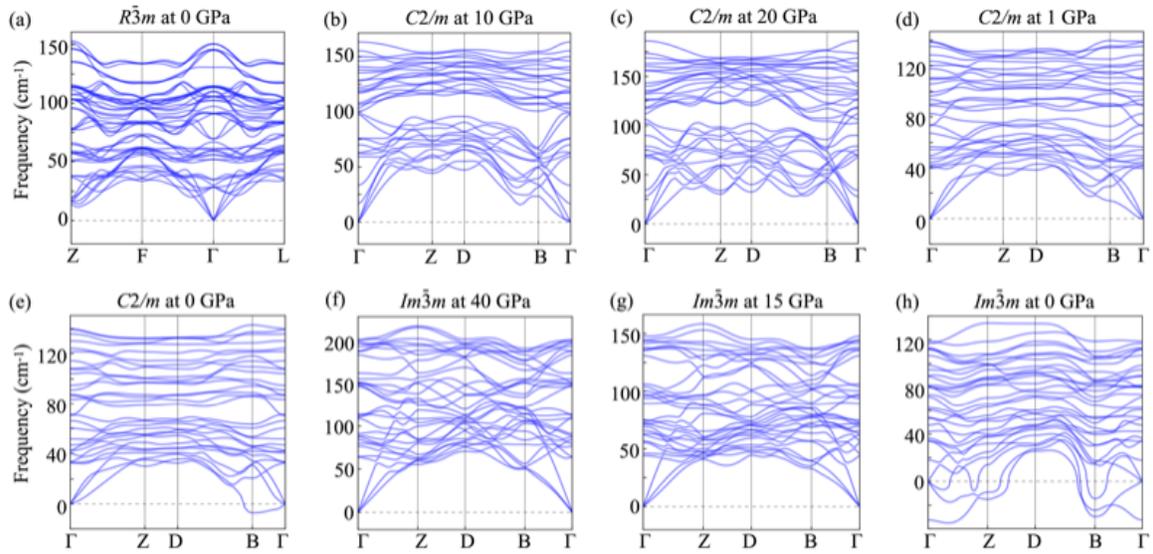

**Figure 6.** Calculated phonon dispersions for $Bi_{0.5}Sb_{1.5}Te_3$ at various pressures. (a) $R\bar{3}m$ at 0 GPa; $C/2m$ at (b) 10 GPa, (c) 20 GPa, (d) 1 GPa (depressurization), and (e) 0 GPa (depressurization); and $Im\bar{3}m$ at (f) 40 GPa, (g) 15 GPa (depressurization), and (h) 0 GPa (depressurization).

Supporting Information

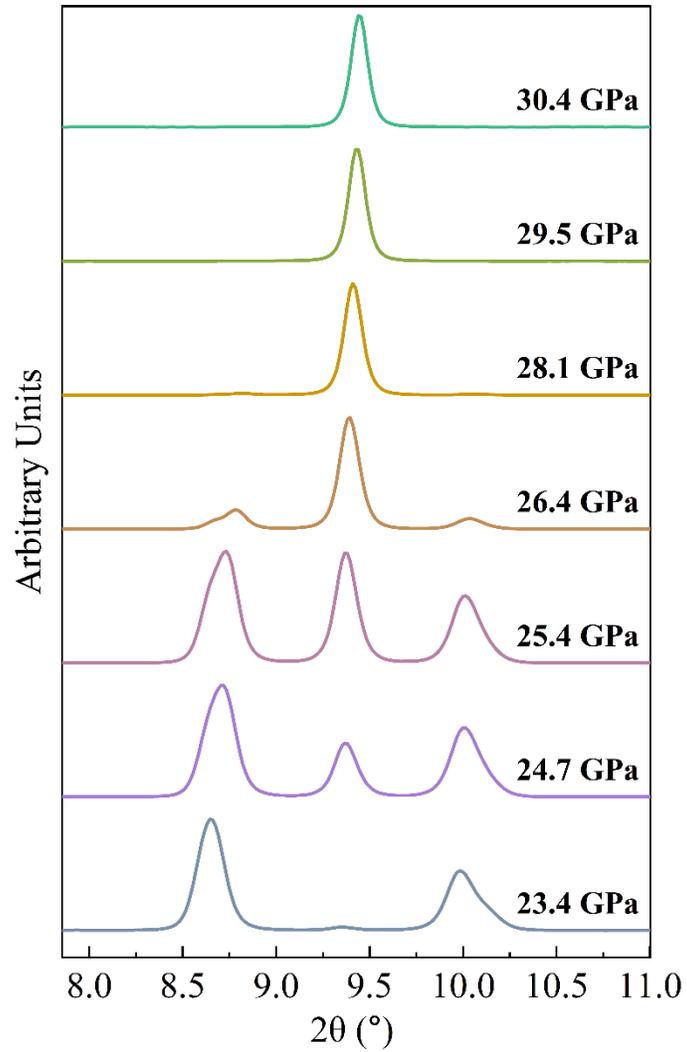

**Fig. S1.** High-pressure X-ray diffraction (XRD) measurements at HPCAT, sector 16, Advanced Photon Source (APS), Argonne National Laboratory (ANL), using a wavelength of 0.4066 Å, showing phase coexistence from 23.4 GPa to 28.1 GPa.

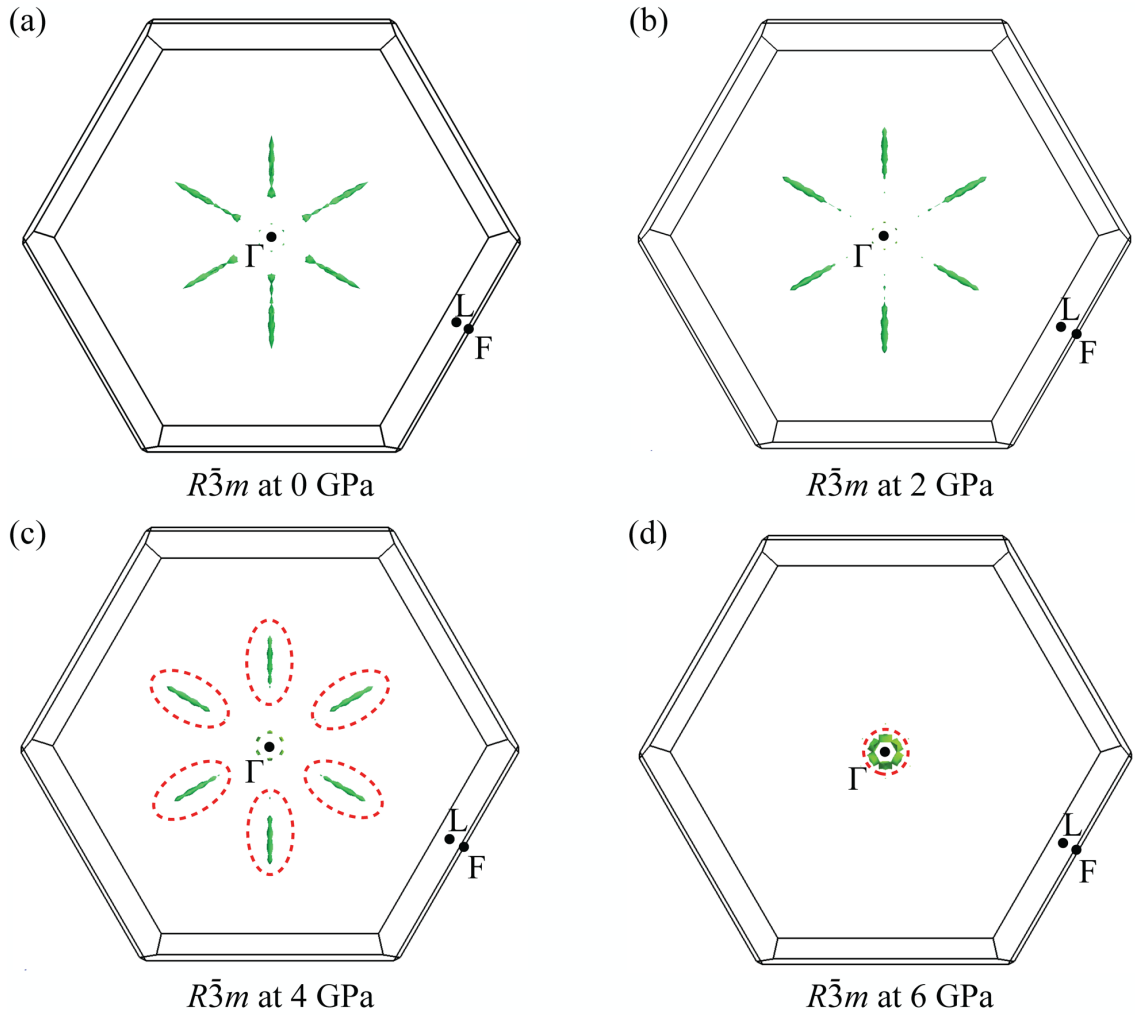

**Fig. S2.** Calculated Fermi surface (FS) for the $R\bar{3}m$ BST structure at (a) 0, (b) 2, (c) 4, and (d) 6 GPa. The areas where the FS topology changes are marked by red dashed ellipses in (c) and the red dashed circle in (d).

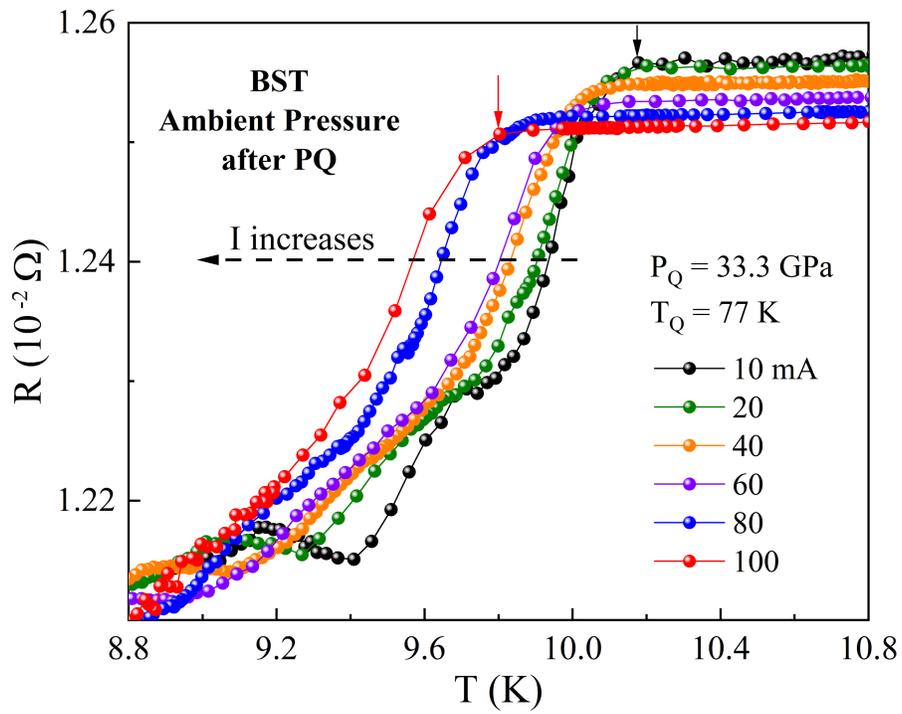

**Fig. S3.** Current effect testing of the retained superconducting phase in $Bi_{0.5}Sb_{1.5}Te_3$ at ambient pressure after pressure quench from $P_Q$ = 33.3 GPa and $T_Q$ = 77 K. The onset $T_c$ decreased from 10.2 K with applied current I = 10 mA to 9.8 K with I = 100 mA, consistent with the superconducting nature of this transition.

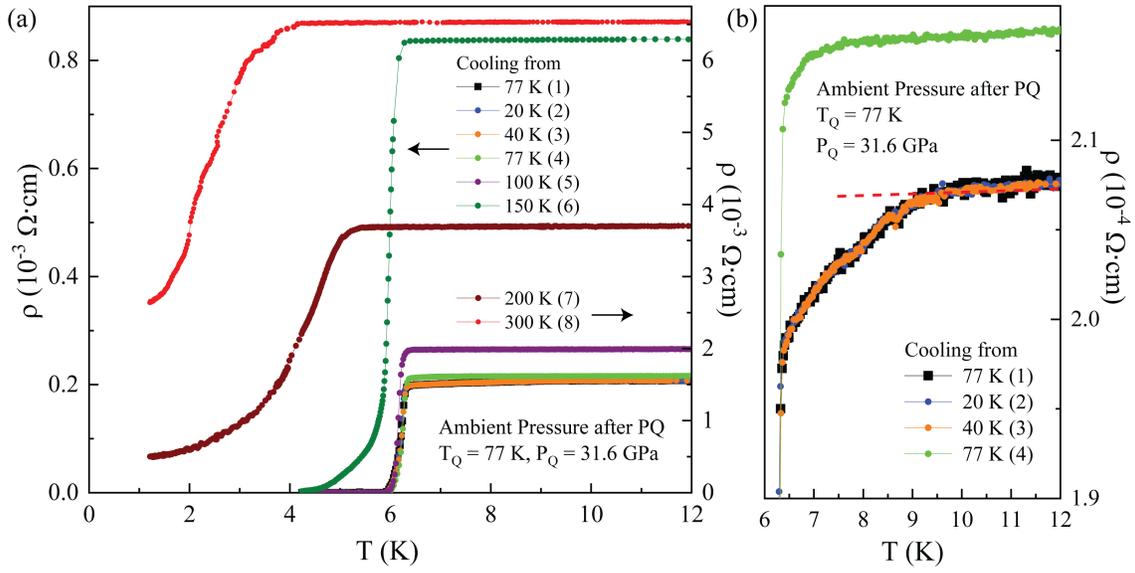

**Fig. S4.** Thermal stability test of the retained superconducting phase in $Bi_{0.5}Sb_{1.5}Te_3$ at ambient pressure after pressure quench. (a) Temperature-dependent resistivity after pressure quenching at 77 K from 31.6 GPa under different thermal cycles up to room temperature. The numbers in parentheses indicate the sequence of thermal cycles. (b) Magnified view of the first four thermal cycles up to 77 K shown in (a), indicating that the higher transition with $T_c \sim 10$ K was quickly suppressed as the sample was warmed up to higher temperatures. The red dashed line is a guide for the eye and shows that a $T_c$ of 10 K was retained at ambient pressure and survived under thermal cycles below 77 K.

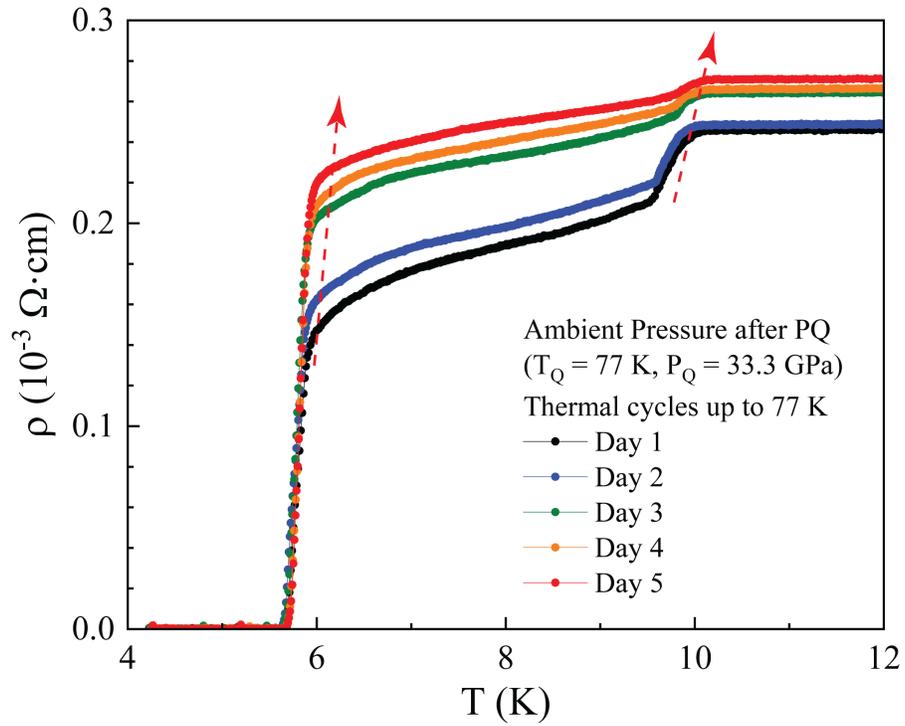

**Fig. S5.** Temporal stability test of the retained superconducting phase in $Bi_{0.5}Sb_{1.5}Te_3$ at ambient pressure after pressure quench. Temperature-dependent resistivity after pressure quenching at 77 K from 33.3 GPa under different thermal cycles up to 77 K measured at different times. Red dashed arrows are guides for the eye and show that there were small temperature increases in the superconducting transitions over time.

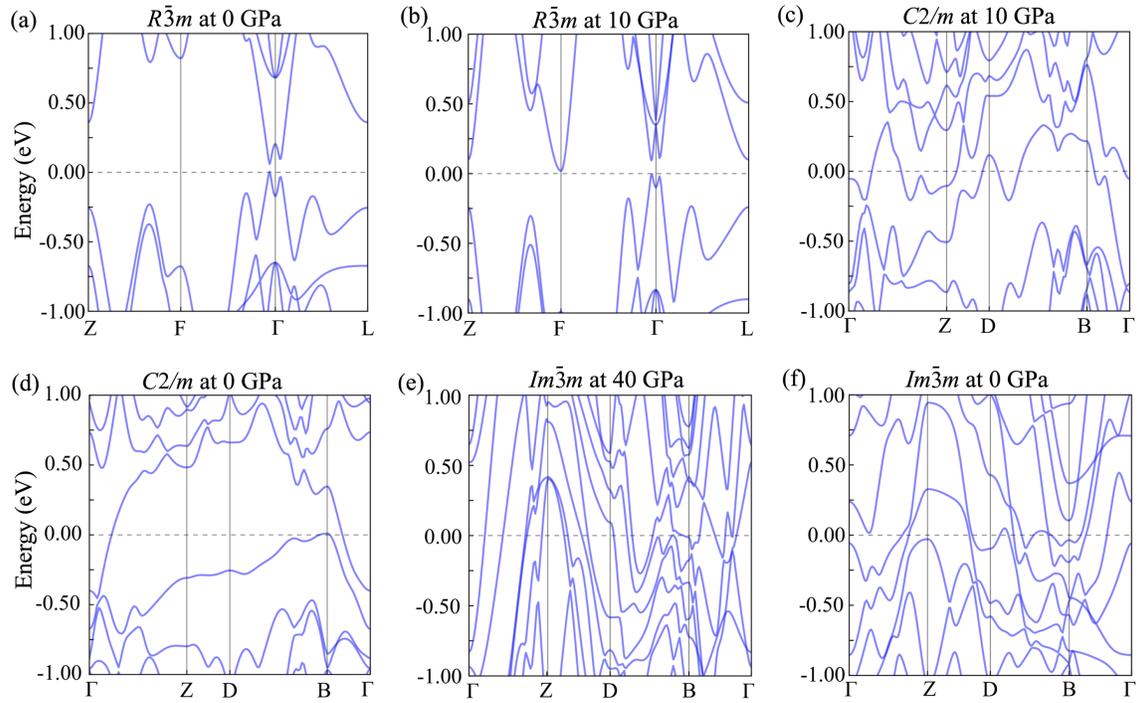

**Fig. S6.** Calculated electronic band structures for Bi$_{0.5}$Sb$_{1.5}$Te$_3$ at various pressures. $R\bar{3}m$ at (a) 0 GPa and (b) 10 GPa; $C/2m$ at (c) 10 GPa and (d) 0 GPa (depressurization); and $Im\bar{3}m$ at (e) 40 GPa and (f) 0 GPa (depressurization).

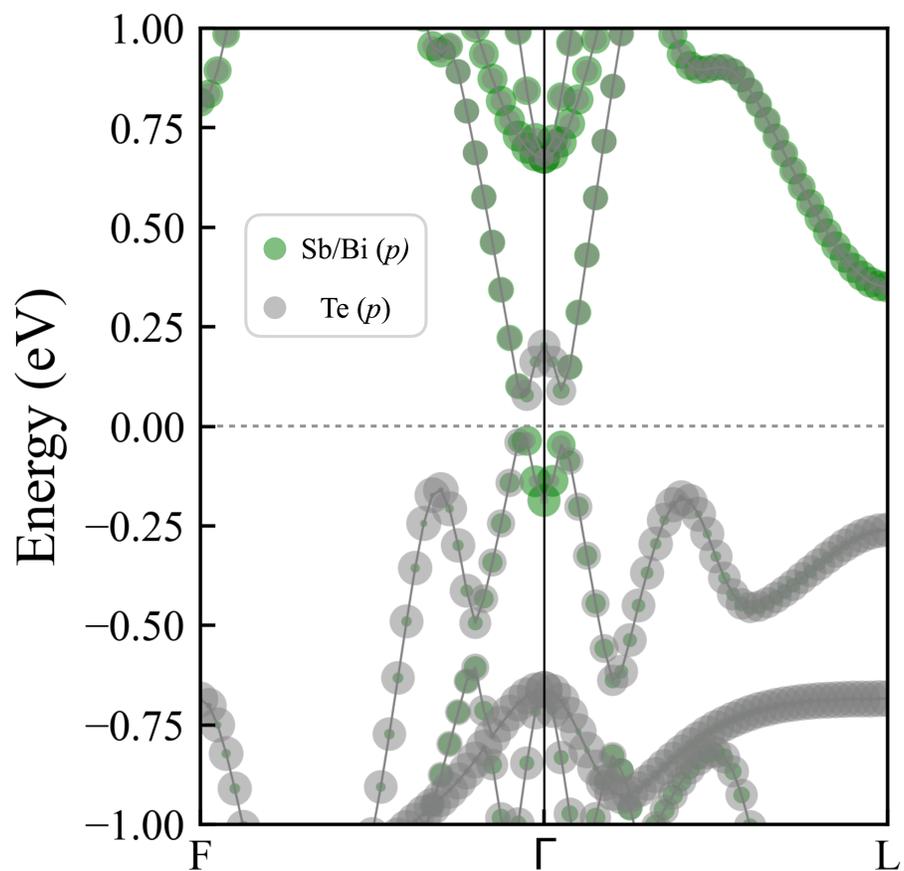

**Fig. S7.** Calculated projected band structure along the K-path from F to Γ to L for the $R\bar{3}m$ crystal structure at 0 GPa. The contributions of the Sb/Bi-*p* orbitals and Te-*p* orbitals are illustrated using green and gray circles of corresponding sizes, respectively. The Fermi level is defined as zero.

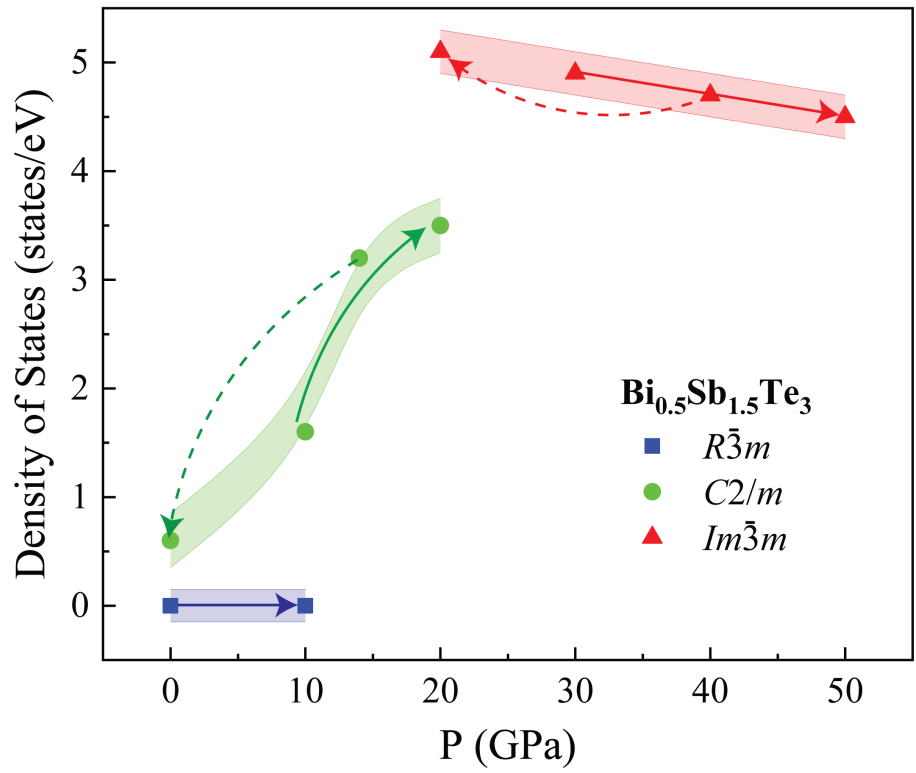

**Fig. S8.** Density of states (DOS) at the Fermi level ($E_F$) as a function of pressure for different structures of $Bi_{0.5}Sb_{1.5}Te_3$. Solid and dashed arrows represent pressurization and depressurization processes, respectively.

**Table S1.** Calculated superconducting $T_c$s under different pressures during pressurization (Press.) and depressurization (Depress.).

| Phase | Pressure (Press./Depress.) | $T_c$ |
|---|---|---|
| $R\bar{3}m$ | 9 GPa (Press.) | 1.3 K |
| $C2/m$ | 10 GPa (Press.) | 1.7 K |
| | 14 GPa (Press.) | 2.3 K |
| | 20 GPa (Press.) | 2.6 K |
| | 1 GPa (Depress.) | 1.2 K |
| $Im\bar{3}m$ | 40 GPa (Press.) | 1.7 K |
| | 20 GPa (Depress.) | 6.0 K |
| | 15 GPa (Depress.) | 7.2 K |